\documentclass[12pt]{article}
\usepackage{graphicx}
\usepackage{color}
\usepackage{amsmath,slashed, amssymb}
\usepackage{fancyhdr}
\usepackage{hyperref}
\usepackage[titletoc]{appendix}
\numberwithin{equation}{section} 

\def\beq{\begin{eqnarray}}
\def\eeq{\end{eqnarray}}
\def\bea{\begin{eqnarray*}}
\def\eea{\end{eqnarray*}}

\def\centeron#1#2{{\setbox0=\hbox{#1}\setbox1=\hbox{#2}\ifdim
\wd1\rangle\wd0\kern.5\wd1\kern-.5\wd0\fi
\copy0\kern-.5\wd0\kern-.5\wd1\copy1\ifdim\wd0\rangle\wd1
\kern.5\wd0\kern-.5\wd1\fi}}
\def\ltap{\;\centeron{\raise.35ex\hbox{$\langle$}}{\lower.65ex\hbox{$\sim$}}\;}
\def\gtap{\;\centeron{\raise.35ex\hbox{$\rangle$}}{\lower.65ex\hbox{$\sim$}}\;}

\newcommand{\newc}{\newcommand}
\newc{\qbar}{{\overline q}}
\newc{\Kahler}{Kahler }
\newc{\deltaGS}{\delta_{\rm GS}}
\begin{document}
\begin{titlepage}
\begin{flushright}
{\large SCIPP 18/02\\
\large ACFI-T17-08 \\}
\end{flushright}

\vskip 1.2cm

\begin{center}

{\LARGE\bf Remarks on the Debye Length and the Topological Susceptibilty in Non-Abelian Gauge Theory}

\vskip 1.4cm

{\large Michael Dine and Di Xu }
\\
\vskip 0.4cm
{\it $^{(a)}$Santa Cruz Institute for Particle Physics and
\\ Department of Physics, University of California at Santa Cruz \\
     Santa Cruz CA 95064  } \\
     ~\\
%{\it $^{(b)}$Amherst Center for Fundamental Interactions, Department of Physics,\\ University of Massachusetts, Amherst, MA 01003
%} \\
\vspace{0.3cm}

\end{center}

\vskip 4pt

%\vskip 1.10cm

\begin{abstract}
We study the Debye mass, $m_D$,  and the topological susceptibility, $\chi$, at high temperatures in non-abelian gauge theory.  Both exhibit, at some order in the
perturbation expansion, infrared sensitivity.  As a result, a perturbative analysis can at best provide an estimate of these quantities, subject
to some uncertainty.  The size of these uncertainties, particularly in the case of $\chi$, has been the subject of some debate.  For the perturbative free energy,
reframing an analysis of Braaten and Pisarski, the estimate and the associated error, can be understood in terms of a {\it Wilsonian} effective action
for the low energy effective three dimensional theory.  This action can be obtained completely from a perturbative calculation, which terminates at a finite order.  This action provides the desired
estimate.  The size of the error follows from dimensional analysis in the low energy theory.
The Debye length computation and its error can be obtained from a similar study of a non-relativistic
effective theory for an adjoint scalar in three dimensions.  $\chi$ requires a four dimensional analysis involving finite temperature instantons, but again the dominant sources of uncertainty are three dimensional, and we provide a procedure
to estimate and an associated error.   This uncertainty is of order $g^4 $ relative to the leading semiclassical
result, and in situations of interest is small.
\end{abstract}

\end{titlepage}

\section{Introduction}

At high temperatures, non-abelian gauge theories undergo a phase transition to an unconfined
phase.  %They become, in some very rough sense, weakly coupled.
The high temperature theory exhibits two mass or length scales.
The first of these is the Debye mass, $m_D \sim g T$, loosely speaking a scale beyond which static electric charges are screened.  The second arises because the theory
at high temperatures and long distances becomes a three (Euclidean) dimensional Yang-Mills theory with coupling
\beq
g_3^2 = g^2 T
\eeq
without matter fields.  %If there are fermions in the theory, their
%contributions at high temperatures to long distance, equilibrium correlation functions is suppressed. 
  The second scale is the mass gap of this theory, on dimensional grounds, $m_{mag} = g_3^2 T$.
  
But the high temperature theory is not exactly a weakly coupled theory.
If one attempts to formulate a
perturbation theory, quantities like the free energy, and even short distance Green's functions, suffer severe infrared divergences, and can be calculated,
at best at low orders.  This is usually described by saying that these quantities are logarithmically divergent at some order, diverging with an additional power of momentum at each further order.  Assuming an infrared cutoff
\beq
m_{mag}^2 = a{g^4 T^2} = a g^4 T^2,
\eeq
each additional order makes a comparable contribution.  For the free energy, divergences first arise at order $g^6$ (four loop order).  
%For the $\vec A$ two point function, this occurs at order $g^4$ (which is consistent with the argument for a magnetic mass of this order).  For the $A_4$ two point function at zero Matsubara frequency and momentum $p^2 \ge 0$, infrared divergences appear only at order $g^8$.
%The $A_4$ two point function, at non-zero Euclidean momenta, exhibits such divergences at order $g_3^4$.
Gauge invariant Greens functions,
like $\langle F^2(x) F^2(0) \rangle $ similarly exhibit such divergences at high enough order.

All of this arises because the theory, at high temperatures, is a three dimensional theory, with a dimensionful coupling $g_3^2 = g^2 T$.
At best, one can hope for a perturbation expansion valid for short distances or high momenta, $g_3^2 r \ll 1~; g_3^2/\vert p \vert \ll 1$.
But loop corrections, even in these limits, are dominated, at sufficiently high order, by low momenta, leading to a breakdown
of weak coupling.

On the other hand, if one computes a {\it Wilsonian} effective action for the three dimensional theory, integrating out momenta between
scales 
\beq
{1 \over \epsilon} g_3^2 < k < \Lambda \sim T
\eeq
one should have a valid expansion in powers of $\epsilon$.\footnote{Our discussion can be viewed as a reframing of an analysis
of Braaten and Pisarski\cite{braatenpisarski}.}  The remaining contributions to physical quantities, Greens functions, and the like
must be obtained from a fully non-peturbative analysis of the strongly coupled three dimensional theory.  This suggests that quantities such
as the free energy can be calculated as a sum of two parts:  the perturbative, Wilsonian contribution, which can be obtained reliably,
and the non-perturbative contribution.   This latter is typically, on dimensional grounds, a power of $g_3^2$ times
an unknown, dimensionless number.   Assuming that the dimensionless number is of order one,
this means that, with a straightforward (if possibly challenging) perturbative
computation, one can obtain an estimate of such quantities, accompanied by an error estimate,
of irreducible size.

Applied to the free energy, as we will explain in section \ref{freeenergyexpansion}, this means that one can reliably compute through order $g^4$.
At order $g^6$, there is a contribution which, again, can be reliably extracted proportional to the log of the ultraviolet cutoff,
and a contribution without a log which cannot be obtained perturbatively.  This non-perturbative contribution represents the irreducilble
uncertainty.

For the Debye length, we will see that there is a similar story.
The existence of a mass -- or correlation length -- for $A_4$ is well known.
At finite temperature, one does not have the full $O(4)$ (in Euclidean space) symmetry.  
At one loop, if one calculates the vacuum polarization tensor, gauge invariance and the remaining
$O(3)$ symmetry are enough to insure vanishing of $\Pi_{ij}$ as $\vec q \rightarrow 0$.  However, this is
not the case for $\Pi_{44}$ at one loop.  If $q_0$
is the discrete frequency of the finite temperature theory,  one finds that for $q_0=0$,  as $\vec q \rightarrow 0$, 
\beq
\Pi^{44}(0,\vec q) \rightarrow m_D^2 \equiv  g^2 T^2 (N + 3 N_f)
\eeq 
In coordinate space, this mass for the $A_4$ field translates into a characteristic length scale.   The $A_4$ Greens function, in
leading order and at large distances, is given by
%\beq
%D^{44}(y) = {1 \over (2 \pi)^3} \int d^3 p {e^{i \vec p \cdot \vec y} \over  p^2 + m_D^2}.
%\eeq
%The integrand has a pole at $p^2 = -m_D^2$, the on shell point, and this can be used to perform the integration as a contour integral.
%This integral can be written in the form:
%\beq
%D_{44}(y) = {1 \over 2 \pi^2 i} \int_{-\infty}^\infty  dp{ p e^{ipy} \over y ( p^2 + m_D^2)}
%\eeq
%It can then be treated as a contour integral; deforming into the upper half plane.  The contour encloses a pole at $q = im_D$,
%This yields
\beq
D^{44}(0,\vec y)= {1 \over 4 \pi \vert  \vec y\vert } e^{-m_D y}.
\label{lowestordergreensfunction}
\eeq

As a result of these considerations, there is a scale, of order $\sqrt{g_3 T} \ll T$ (for small $g_3$), at which one has a three dimensional gauge theory with an adjoint
scalar, $\phi$, of mass $\mu = m_D$.
Corrections to the Debye length have been considered in the literature\cite{rebhan,arnoldyaffe}.  We will consider them
from two points of view.  We'll first examine
the direct computation of $D^{44}(\vec y)$ in perturbation theory.  We'll see that one can obtain a reliable estimate of the Greens
function at distances parameterically large compared to $\mu^{-1}$ by a factor $(g_3^2 \log(\mu/g_3^2))^{-1}$.
Beyond this scale, the computation of the Greens function, order by order in the perturbation expansion, is not under control.

But as explained in \cite{arnoldyaffe} and we elaborate further here, it is possible to define a gauge-invariant, non-perturbative Debye mass which controls the very large $r$ behavior of the Greens function.  As we explain, viewing the three dimensional
theory as a {\it Minkowski} theory with an adjoint scalar, the theory has a $Z_2$ symmetry.
\footnote{These are statements about QCD, in the absence of weak interactions.  As noted in \cite{arnoldyaffe}, in theories which are not vector-like,
the low energy theory may not respect the $Z_2$.  In this case, strictly speaking, there is no sharp definition of the Debye mass.
However the $Z_2$ breaking is often highly suppressed.  For example, in non-vector-like theories without
scalars, there is still a $Z_2$, related to $CP$ in the four dimensional theory.  As a result, there is often a range of (large) distances
where correlators invariant under the approximate $Z_2$ exhibit a rapid exponential falloff, even if, at extremely large distances, the falloff
is power law.}
  The mass of the lightest
$Z_2$ odd state controls the Euclidean large distance behavior of Green's functions of $Z_2$-odd operators, and
is naturally defied as the Debye mass.  One can obtain an estimate of this mass by studying the divergence structure of the perturbation theory, noting that there is logarithmic infrared sensitivity at one loop in the computation of this mass.  But the
structure of the perturbation expansion is complicated, with a variety of infrared singularities at higher orders.
Instead, one can also use the Wilsonian language, as for the free energy, applied to a suitable non-relativistic effective action.
Here there is, again, an ultraviolet divergence (the cutoff for the low energy theory is now $\mu = m_D$).
Again, one obtains an {\it estimate} of the Debye length, as well as an irreducible, perturbative uncertainty.

But while providing, perhaps, a different language, for the free energy and the Debye length, this serves simply to confirm longstanding
results.  But our particular interest is in the nature of the semiclassical estimate of $\theta$-dependent effects
at finite temperature, interesting in themselves and relevant to the problem of axion cosmology.
Here the object of interest is the free energy
as a function of $\theta$ and $T$,  $F(\theta,T)$.  Much of the literature, particularly the lattice literature,
focuses on the
topological susceptibility
\beq
\chi(T) \equiv {\partial^2 F \over \partial \theta^2}.
\eeq
 The leading term as a function of $g^2$ can be computed in the dilute gas approximation, and is known\cite{gpy}.  At some order, one expects infrared divergences to arise as in the computation of the free energy in ordinary perturbation theory.  One of the goals of this paper is to determine the order of the corrections to the leading semiclassical approximation
at which theses divergences arise.  We will seek to determine the nature of these divergences, in order to assess the uncertainties in the standard computation.
We will examine the large distance behavior of Feynman diagrams in an instanton background, determining the
order of the expansion in $g_3^2$ about the classical solution at which first logarithmic divergences and then power law divergences
arise.  We will interpret the log divergence, as for the perturbative free energy, as a term in a Wilsonian action involving the log
of the cutoff ($\Lambda =T$) .   This will, again, permit us to make an estimate of $\chi$ and to determine an irreducible uncertainty
in the semiclassical computation.

Apart from the intrinsic interest of finite temperature gauge theory, our work has been motivated, in part, by arguments in the literature for
large corrections to the semiclassical computation of $\chi$ ($F(\theta, T)$.  In particular, it has been asserted that  at one loop there is a correction to the instanton action which is (fractionally) of order order $g \log g$ rather than $g^2$ and that at temperatures of interest it is numerically of order one.  This effect is exponentiated in $\chi$, and as such
could lead to an uncertainty of orders of magnitude.  This claim has been used, in turn, to argue that lattice computations are essential to determine the behavior of hypothetical axions in the early universe.  Such computations have yielded values of $\chi$ varying by orders of magnitude at relevant temperatures, both from each other and from the leading semiclassical result.
\cite{bonatia,bonatib,borsanyia,borsanyinature,villadoro}.

The basis for these concerns is the assertion that the corrections to the Debye length which we have described above
are large\cite{rebhan,arnoldyaffe} and that the Debye length acts as an infrared cutoff on the instanton size in the dilute gas approximation.  As explained in
\cite{gpy}, the Debye {\it mass term in the effective lagrangian} does provide an infrared cutoff, but as noted in \cite{dinedraperinstantons}, the actual cutoff in the instanton scale size in the computation of $\chi$ is $T^{-1}$.  As a result, as we will explain further in this paper, uncertainties associated with the cutoff on the $\rho$ integration are small.

Still, the $\theta$-dependence of the free energy will, at some point, exhibit infrared sensitivity.
We will demonstrate that the expansion for the topological susceptibility,  $\chi$
exhibits infrared divergences at lower order than that for the perturbative free energy:
\beq
\chi(T) =a T^4 (1 + b g^2 + c g^4 (\log(g^2) + C).
\eeq
This follows from the explicit form of the instanton, and in particular its large distance behavior.

This paper is organized as follows.  In section \ref{freeenergyexpansion}, we review the high temperature behavior of the free energy in QCD.
In particular, we note that the ``infrared" log is also an ultraviolet logarithm from the point of view of the three dimensional
effective theory, and that the coefficient of this logarithm can be reliably calculated.  As a result, there is a well-justified computation of the free
energy, to a fixed order in $g^2$, whose error can be reliably estimated.

In section \ref{largedistanceg}, we consider the perturbative calculation of the $\phi$ ($A_4$) Green's function, with a focus on the large distance, coordinate
space behavior.   After reviewing the leading computation, we consider broad classes of Feynman diagrams.  These become progressively
more singular at large distances, and large, infinite sets are actually infrared divergent.  Assuming a cutoff $m_{mag}$, one can establish
that the leading correction dominates, up to some maximum distance.

In section \ref{debyemassdefinition}, we explain that, with the assumption the three dimensional theory is gapped, there is a well-defined notion
of the Debye mass, and we provide a definition.  Then, in section \ref{debyemasscalculation}, we turn to the question of to what 
extent we can estimate this mass, i.e. to what order in the perturbation expansion, and with what level of uncertainty.  Using the language of Non-Relativistic Effective Theory (NRET), we will argue
that this computation is robust.

In section \ref{chicalculation}, we turn to the question of the calculation of the topological susceptibility.  We review some features
of the finite temperature instanton computation,
explaining that, at low orders in the semiclassical computation, the dominant instanton scales, $\rho$, are of order $T^{-1}$.  Then we ask
about the appearance of infrared divergences in this computation.  We work in coordinate space; noting that at large distances, the instanton solution falls off rapidly, so that for purposes of isolating the infrared divergence, the modifications of the
relevant Greens functions from their tree level forms are small.  Essentially, we are able to treat the background instanton as a perturbation.  We note that individual Feynman diagrams are actually divergent already at two loops, but gauge invariance
requires that these divergences cancel, and the leading infrared divergence occurs at three
loops (order $g^4$).  Again, we argue that the computation of the logarithmic term at three lops is robust, and gives us both an estimate of the size of $\chi$ and of irreducible uncertainties.

Finally, in section \ref{conclusions}, we consider the implications of these observations for some physical problems.  We focus
on the calculation of the finite temperature axion mass, which is proportional
to $\chi$.  We note, again, that the cutoff on the instanton scale size
integration is of order $T^{-1}$, but stress that this
distance is not only parameterically small compared to the Debye length, but it is {\it even smaller} than the scale at which the
large corrections to the Debye come into play.  We note that the infrared divergence at order $g^4$ implies an error in the computation
of the susceptibility at the $1\%$ level or better, which is not significant for the calculation of the axion energy 
density.
%We mention some connections to other problems of the finite temperature theory as well.

\section{ $g^2$ Expansion of the Free Energy}
\label{freeenergyexpansion}

In this section, we consider the expansion of the free energy in perturbation theory in powers of $g^2$.  For the perturbative free energy, ignoring, at first, the adjoint scalar
(i.e. $A_4$),  there is formally an expansion of the form:
\beq
F(T) = T^4 \sum_{n=0}^\infty a_n g^{2n}
\eeq
but this expansion breaks down due to infrared divergences at a certain order.  This can be understood in terms of the behavior of the three dimensional theory, with effective coupling $g_3^2 = g^2 T$, and an ultraviolet cutoff $\Lambda \sim T$.  Let's first ignore the adjoint scalar $\phi$ and consider the computation of a Wilsonian effective action in the pure gauge theory, integrating out physics between scales $\Lambda$ and $\epsilon \Lambda$, where ${g_3^2 \over \epsilon \Lambda} \ll 1$.  Then in the Wilsonian action, one obtains for the vacuum energy (coefficient of the unit operator) a series:
\beq
E_0 = a\Lambda^3 (1 + {\cal O}(\epsilon)) + b g_3^2 \Lambda^2(1 + {\cal O}(\epsilon)) + c g_3^4 \Lambda(1 + {\cal O}(\epsilon)) + d g_3^6 \log(\epsilon) .
\label{3de0}
\eeq
In individual Feynman diagrams, the logarithmic behavior is readily identified by power counting.   Higher order terms in the expansion are suppressed by powers of
$g_3^2/\Lambda$.  This is characteristic of a superrenormalizable theory.  The computation of the Wilsonian action
terminates at some power of the coupling.

From the requirement ${g_3^2 \over\epsilon\Lambda}\ll 1$, 
we have $1 \gg \epsilon \gg g^2(T)$.  So thinking of $\epsilon$ as several times $g^2$, say $\epsilon = A g^2$, where
$
1 \ll A \ll {1 \over g^2},$
we can reliably say that the Wilsonian action includes a contribution to the coefficient of the unit operator (ground state energy) of the form of equation \ref{3de0}.  In the low energy, cutoff theory, there will be a contribution to
the energy of size $g_3^6 \log(A)$, which requires non-perturbative evaluation.

Including the adjoint field in the analysis, we might expect an expansion in $\mu^2$ and $g_3^2$.  However, already at one loop, the expansion is actually  an expansion in $\mu$; at one loop order, there is a contribution behaving as $\mu^3 T \sim g^3 T^3$\cite{kisslinger}.  Higher orders yield more complicated dependence on $\mu$.

Returning to the infrared perspective we discussed in the introduction, we saw that, beyond the $g^6 \log g$ term, there
are an infinite number of perturbative contributions which are nominally of order $g^6$.  From the perspective
of the three dimensional effective theory, assuming that the vacuum energy is well-defined, after integrating out
modes well above $g_3^2$, any further contribution necessarily scales as $g_3^6$.  Moreover, these contributions
are not perturbatively accessible.  Formally, we can attempt to tame the infrared divergences by various strategies.
Apart from introducing a magnetic mass or simply cutting off momentum integrals at that scale, we can resum the
contributions to the propagator, yielding $1/(g_3^2 k)$ behavior at small $k$.  But it is easy to check that all diagrams
beyond four loop order are of order $g_3^6$.

\section{Perturbative Computation of the Greens function at Large Distances and its Limitations}
\label{largedistanceg}

In the next section we will see that from knowledge of the three dimensional {\it Minkowski} theory we can
obtain information about the large distance behavior of the Euclidean theory.  We first consider the direct computation
of the Green's function in coordinate space, and then move on to non-perturbative considerations.

At tree level, the Euclidean Greens function in coordinate space can be evaluated by Fourier transforming
the momentum space expression.   Performing the angular integrals, the remaining momentum integral (integral over $p$)
can be treated as an integral in the complex plane.  Deforming into, say, the upper half plane one picks up the pole at $p = i \mu$.
This corresponds to the on shell point in the Minkowski description.

As one works to higher order in perturbation theory, the self energy, $\Sigma$, as we will shortly see, has a branch cut starting at $p = i \mu$.
So now for the Greens function, deforming the contour, one encircles the branch cut.  Calling the new integration variable $\delta$, the integral
involves a factor $e^{-(m+\delta) r}$.  For large $r$, it is dominated by $\delta \sim r^{-1}$.  In the Minkowski language, this corresponds to
$\phi$ being off shell by an amount of order $\mu \delta \sim {\mu \over r}$.  So in {\it momentum} space, we are interested in $\Sigma(p)$
for $p^2 = \mu^2 - 2 \mu \delta$, with $\delta$ small.  
We want to ask:  how small can $\delta$ be and still yield a reliable estimate.

Consider, first, the one-loop contribution to the self-energy, $\Sigma(p)$ (figure \ref{oneloopsigma}.  Work slightly off shell:
\begin{figure}[t!]
\includegraphics[width=10cm]{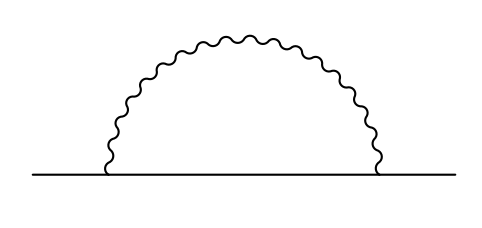}
\caption{The leading contribution to the $\phi$ self-energy, $\Sigma(p)$.} 
\label{oneloopsigma}
\end{figure}
\beq
p^2 = \mu^2 - 2 \mu \delta.
\eeq
Then 
\beq
\Sigma = N g^2 \int {d^3 k \over (2 \pi)^3} {4\mu^2 \over k^2+i \epsilon}{1 \over 2\mu (-\delta + k^0)+ i \epsilon}
\label{sigmaoneloop}
\eeq
$$~~~~= {N g^2 \over  \pi} \mu \log(\mu/\delta)$$
Adding $\alpha {k^\mu k^\nu \over k^2}$ to the propagator, it is easy to check that the logarithmic term in this expression is gauge invariant.

%Before considering the Fourier transform, we examine higher order corrections to $\Sigma$, asking for what range of
%\delta$ these are small.
As we will now show, successive orders in the expansion of $\Sigma$ are progressively
more singular in $\delta$.  Moreover, for fixed $\delta$, one encounters actual infrared divergences at two loop
order and beyond.

We can consider several classes of higher order perturbative corrections to $\Sigma(p)$ to illustrate the behavior
of the perturbation expansion for small $\delta$ and to determine where it breaks down.  There are three issues:
\begin{enumerate}
\item  Singular behavior for small $\delta$
\item  Actual infrared divergences
\item  Divergent series for some value of $\delta$ and plausible infrared cutoff.
\end{enumerate}
  One class of diagrams involves ``rainbows"
of gluons emitted by $\phi$ (figure \ref{rainbows}).  For these, a typical contribution is of the form:
\begin{figure}[t!]
\includegraphics[width=10cm]{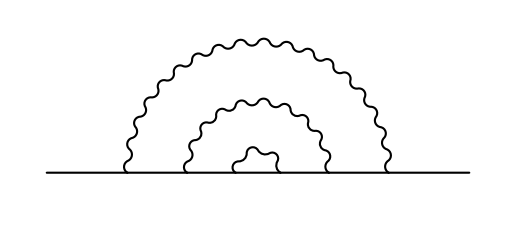}
\caption{One class of diagrams singular in the limit $\delta \rightarrow 0$.} 
\label{rainbows}
\end{figure} 
\beq
g_3^{2n}\int {d^3 k_1 \dots d^3 k_n \over (2 \pi)^{3n} k_1^2 \dots k_n^2} {1 \over 2\mu  (k_1 + \delta)}{ 1 \over 2\mu(k_1 +k_2+ \delta)}
\dots  {1 \over 2\mu(k_1 +k_2+ \delta)}{1 \over 2\mu(k_1 + \delta)  }
\eeq
$$\sim g^{2n} \delta^{-(n-1)}.$$
If we restrict
\beq
\delta \ge {N g^2 \over  \pi} \mu \log(\mu/\delta),
\label{deltacondition}
\eeq then the perturbation series would appear to
be an expansion in powers of $1/\log(\delta)$.

%At next order, there will be a  correction proportional to ${g^2 \over \delta} = {1 \over \log (\mu/g_3^2)}.$.
%If these were the only
%contributions singular for small $\delta$, we would have an expansion
%in powers of $1 \over \log (\mu/g_3^2)$.  But we have to consider another class of contributions.

\begin{figure}[t!]
\includegraphics[width=10cm]{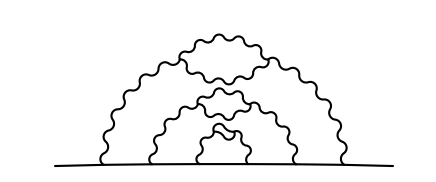}
\caption{A second class of diagrams singular in the limit $\delta \rightarrow 0$, with actual infrared divergences.} 
\label{rainbowloops}
\end{figure}

But other classes of diagrams leads to a stricter requirement.
The first consists of ``rainbows" as above, but each with a one loop vacuum polarization correction
on the gluon line (figure \ref{rainbowloops}).
Such diagrams, by
power counting, have an actual infrared divergences, and are also more singular as $g_3^4/\delta^2$ for each
additional loop.  In other words, assuming an infrared cutoff of order $g_3^2$, each additional loop comes with a factor: 
\beq 
 {g_3^4 \mu \log(m_{mag}/\mu) \over \delta^2}.
\eeq
%So this leads to the more stringent requirement,
%\beq
%\delta \gg  {N g^2 \over  \pi} \mu \log(\mu/\delta).
%\eeq
So, for $\delta$ satisfying the condition \ref{deltacondition}, these contributions are all nominally of similar size,
each suppressed by
$\log(\delta) \sim \log(\mu/g_3^2)$ relative to the leading contribution.  %There is also a numerical suppression factor which grows with the
%order of the perturbation expansion.

\begin{figure}[t!]
\includegraphics[width=10cm]{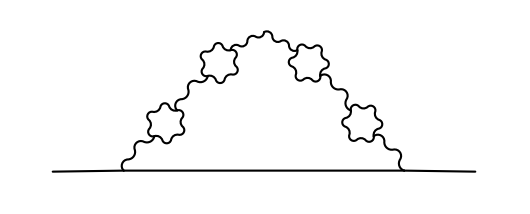}
\caption{One class of diagrams singular in the limit $\delta \rightarrow 0$.} 
\label{dressedgluon}
\end{figure}

A third class of diagrams involve propagator corrections to the gluon line of the one loop contribution (figure \ref{dressedgluon}).
With $n$ one-loop corrections to the gluon line, these are infrared divergent, behaving as
\beq
\delta \Sigma^{(n)} \sim (g_3^2)^{n+1} \int {d^3 k \over k^2 k^n} \sim {g_3^2 \mu \over n \delta} {(g_3^2)^n \over m_{mag}^{(n)}}
\eeq
These diagrams are individually suppressed at large $\delta$, only by a single power of $\delta$; the series $\sum_{n} {1 \over n}$
is log divergent.  Assuming that this sum is of order $\log(\mu/m_{mag})$, this requires,
again, that $\delta \gg g_3^2 \log(\mu/m_{mag})$.

\begin{figure}[t!]
\includegraphics[width=10cm]{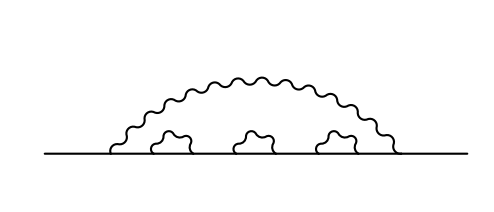}
\caption{One class of diagrams singular in the limit $\delta \rightarrow 0$.} 
\label{dressedphipropagator}
\end{figure}

A similar requirement arises from the class of diagrams in figure \ref{dressedphipropagator}.
These behave, with $n$ one-loop corrections to the $\phi$ propagator, as
\beq
\delta \Sigma^{(n)} \sim g_3^2 \int {d^3 k \over k^2 (k+\delta)^n}(g_3^2 \log (\delta/\mu))^n \sim {g_3^2 \over n} (g_3^2 \log (\delta/\mu))^n.
\eeq
So at each order in $n$, we have a comparable contribution, suppressed by a logarithm relative to our leading contribution.  
The sum would appear poorly behaved.  But if we sum before integration, we obtain a correction to the leading contribution down
by a logarithm, in other words, of order $g_3^2$.

%Higher loop corrections, however, would appear to be of order
%\beq
%\Delta \Sigma^{(n)} \sim {1 \over n} 
%\eeq

%So the perturbative expansion breaks down with corrections of order $g_3^2$.  This non-calculable term is consistent with expectations from confinement, equation \ref{confinement}.

Of course, there are many other diagrams -- mixtures of these various types and other topological classes altogether.
The main lesson is that, {\it at best}, one can calculate for $\delta \gg \delta_0= {g_3^2 N \over 2 \pi} \log(\mu/g_3^2)$.
With this understanding of the behavior of $\Sigma$, let's consider the fourier transform of the  propagator in Euclidean space,
\beq
G(r) = \int {d^p k \over (2 \pi)^3} e^{i \vec p \cdot \vec x} %{1 \over k^2} %{1 \over 2 \mu k^0 + \mu \delta - \Sigma(p)}
{1 \over k^2 +\mu^2 -\Sigma(p)}
 \eeq
 Performing the angular integrals leaves:
 \beq
 G(r) = \int_{-\infty}^\infty {dp p \over (2 \pi)^2}{1 \over i r} {e^{i p r} \over p^2 + \mu^2 -\Sigma(p)}. 
 \eeq
 The integrand has  branch cuts starting at $p =\pm i\mu^2$.  Treating as a contour integral and deforming
 so as to encircle the branch cut in the upper half plane, the integral becomes:
 \beq
 G(r) = \int_{\mu}^\infty {d\delta \rho \over (2 \pi)^2}{1 \over  r} e^{-\delta r}{\rm Disc}{1 \over 2\delta \mu +\Sigma(\delta)}. 
 \eeq
 The main noteworthy feature here is that for large $r$, $\delta \sim {1 \over r}$.  So for sufficiently large $r$,
 \beq
 r \approx (\log ({\mu \over g_3^2}) g_3)^{-1}
 \eeq
 one has lost control of the expansion of $\Sigma(p)$, and the computation of the propagator has broken down.   Still, parameterically
 in $g_3$, there is a large range of distance where the propagator can be reliably estimated.
 
 In the following sections, we will see that general arguments establish that the falloff of the propagator is that of a massive
 scalar field.  The corrections to the mass can be estimated with a definite error.  The results are consistent with our estimates
 above, in that the propagator is exponentially modified from its perturbative form at distances $r >\delta_0$, but the error on
 the exponent is controlled.

\section{Defining the Debye Mass}
\label{debyemassdefinition}

At high temperatures, as is well known, four dimensional field theories with massless fields behave as three dimensional systems.
In the case of Abelian gauge theories with light matter, and general non-abelian theories, the field $A_4$ of the four dimensional
theory behaves like a massive field, with mass parameterically less than $T$ by a single power of $g$, the gauge coupling.
In this section, we'll consider a theory with a massive adjoint field, $\phi$, with lagrangian mass parameter $\mu^2$, small compared to some ultraviolet cutoff, and attempt to determine the behavior
of the charged field propagator at large distances.  

In QCD, the resulting effective theory has a $Z_2$ symmetry, arising from the four dimensional symmetry of the Euclidean
theory $A_4 \rightarrow -A_4,~~t \rightarrow -t$.  Calling $A_4 \equiv \phi$ (with corresponding transformations
on the fermions), the symmetry takes $\phi \rightarrow -\phi$ in the three dimensional theory.

To extract properties of the theory at large Euclidean distances, it is helpful to consider the theory continued to three dimensional {\it Minkowski}
space (the argument which follows is familiar in lattice gauge theory and other contexts).  
We expect in the three dimensional system, there will be states odd under the $Z_2$, which are bound states of
$\phi$ and gluons.  Suitable interpolating fields for such states would be:
\beq
\Phi = {\rm Tr}(\phi F_{\mu \nu}^2),
\eeq
and operators with more $F_{\mu \nu}$ and/or $D_\mu$ factors.

%The mass of the lowest state in this channel we would expect to be
%equal to that of the $\phi$ particle calculated in perturbation theory, up to corrections of order $m_{mag}$.  Correspondingly, we would
%expect that the location of the poles in gauge invariant Greens functions would differ from those of $\phi$, in a particular gauge, by amounts of order $m_{mag} \mu$.

If we continue the three dimensional {\it Euclidean}
theory to Minkowski space, we can write a spectral representation of the Green's function:
\beq
G_\Phi(x) = \langle \Phi(x) \Phi(0) \rangle = \int dM \rho(M) D_F(x,M).
\label{spectralrepresentation}
\eeq
In \ref{spectralrepresentation}, $D_F(x,M)$ is the free propagator for a field of mass $M$.  If the spectrum is gapped,
we can write
\beq
\rho(M) = Z \delta (M-M_{phys}) + \theta(M-M_0) f(M)
\eeq
where we will refer to $M_{phys}$ as the {\it physical mass} of the lightest $Z_2$-odd particle, and $M_0 > M_{phys}$.  Then we can continue eqn. \ref{spectralrepresentation} back to
Euclidean space, and (in three dimensions) show that the asymptotic behavior of the propagator is:
\beq
G_\Phi(x) = {Z \over 8 \pi r} e^{-M_{phys} r}
\eeq
for $r = \vert \vec x \vert \gg M_{phys}^{-1}$.  So the spatial falloff of the propagator is determined in terms of $M_{phys}$.  This quantity is gauge invariant.
If we can estimate $M_{phys}$ in some regime where the perturbation expansion is reliable, then, using unitarity, we can continue
to the regime of arbitrarily large distance, with an error in the estimate which we can hope
to control.  $M_{phys}$ we define to be the Debye mass.

%On the other hand, if $\rho$ is a continuous function of $M$, with no gap, the behavior of the propagator is more complicated
%and not determined a priori.  As we will remark shortly, this is the case of the $U(1)$ theory.  With plausible, and rather
%standard assumptions, $\rho(M)$ {\it is gapped} in the non-abelian theory.

\section{Calculating the Debye Mass}
\label{debyemasscalculation}

In this section, we determine how well one can estimate the Debye mass, and compare with the analysis of section
\ref{largedistanceg}.

\subsection{Non-Relativistic Effective Theory (NRET)}

In our discussion of the perturbative free energy, we were able to argue that the {\it infrared} logarithm at four loop order
was robust; it could be understood as a renormalization of the unit operator of the effective three dimensional theory
at low energies.  Given that our focus, for the Debye mass, is also on infrared (long distance) issues, we can ask
whether we can isolate a similar {\it ultraviolet} logarithm.  

%It is natural to ask if this computation is meaningful, in the same way we argued that the
%computation of the $g^6 \log g^2$ contribution to the free energy is robust. After all, the various re-summations we have
%performed are not obviously systematic (or even gauge invariant).  To assess this, 
We can consider the problem from the point of view of
non-relativistic effective field theory.  For scalars (see, for example, \cite{schwartzfieldtheory}), one conventionally defines:
\beq
\phi = {1 \over \sqrt{2 m}} e^{-i m v \cdot x} \chi; ~~v^2 =1.
\eeq
%Here we take $m^2 = \mu^2 + \delta \mu$.
The action for $\chi$ is then:
\beq
i \chi^* v_\mu D^\mu \chi .%- \delta \chi^* \chi.
\eeq
%Then there is an extra term in the action, $\delta \chi^\dagger \chi$.
At one loop, there is an ultraviolet and infrared divergent correction to the $\chi$ propagator, similar to the one loop correction
to $\Sigma$ which has been our focus:
\beq
{g_3^2 N \over 2 \pi} \log(\Lambda/m_{mag}).
\eeq
For the non-relativistic theory, the ultraviolet cutoff is the mass, $\mu$.  This divergence is cured by a counterterm for the
operator $\chi^\dagger \chi$; it corresponds to a mass shift in the non-relativistic theory.  Because the operator has dimension
$2$, and $g_3^2$ has dimension one, there is only an ultraviolet divergent correction at one loop.  
 We can again think of the logarithm as a term arising in a Wilsonian effective action
from integrating out high energy modes of the gauge field, in this case, between the ultraviolet cutoff, $\mu$, and a scale 
\beq
\lambda = {1 \over \epsilon} g_3^2.
\eeq
In principle $\epsilon$ is a small number,  $1 \gg \epsilon >> {g_3^2 \over \mu}$.
 
This counterterm
eliminates all $\mu$-dependence in the perturbation expansion.   The theory does contain the parameter $\log \epsilon$,
which can be thought of as an order one number.   Thus the effective theory has only the dimensionful parameter,
$g_3^2$.  So necessarily any further correction to the mass, beyond the counterterm, is a constant times $g_3^2$.
In other words, this argument establishes that
\beq
m^2 = \mu^2 + {g_3^2 N \over 2 \pi} (\log(\mu/g_3^2)  + A).
\eeq

 The results of the NRET analysis can be understood in a more conventional
 Feynman diagram analysis.  Dividing up the integration over the gluon momentum into two regions, one
 with momentum $k > k_0 = {1 \over \epsilon m_{mag}}$, and one with $k < k_0$, we also take the external momentum
 to satisfy
 \beq
 p^2 = \mu^2 + \delta_0 + \delta^\prime; ~~\delta_0 = {N g^2 \over 2 \pi} \log(\mu/k_0).
 \eeq
 where $\delta_0$ represents the lowest order mass shift.  We take the propagator to be the resummed propagator,
 with the one loop contribution to $\Sigma$.  With this choice, all higher loop contributions to $\Sigma$ are dominated by the infrared, and are thus insensitive to $\mu$, apart from an overall factor.  They behave as $(g_3^2 \mu)(g_3^2/ \delta^\prime)^n$.
 The mass shell condition takes the form
 \beq
 \Sigma(\delta^\prime) = \mu \delta^\prime.
 \eeq
 So assuming $\Sigma(\delta^\prime) = g_3^2 \mu A({g_3^2 \over \delta^\prime})$ we have that $\delta^\prime = a g_3^2$
 for some constant $a$.

%The dimensionful scale, $\lambda_{ir}$, is necessarily proportional to $g_3^2$.  But, as we have seen, there are many types
%of infrared effects, and the present one is somewhat subtle, as it involves behavior as one approaches the mass shell.
%In the next subsection, we isolate the leading infrared logarithm (reproducing the result of \cite{rebhan}).

\subsection{Comparison of NRET Analysis to Conventional Perturbation Theory}

The argument based on non relativistic effective theory gives a sharp, principled argument that one can calculate, perturbatively, a correction of order $g_3^2 \mu \log(\mu/g_3^2)$ to the mass of $\phi$, while the remaining correction is of the form $A g_3^2$. 
Note that this is in accord with our discussion of the previous section.  We an compute, in conventional perturbation theory,
the Green's function out to distances of order $r \ll {1 \over g_3^2 \log ( \mu/g_3^2)}$.  Beyond this distance scale, the
corrections quickly become large compared to one.  The non-perturbative analysis gives us a reliable estimate of the 
Green's function to scales of order $ r\sim g_3^{-2}$.  Note that both of these scales are parameterically large compared to $\mu^{-1}$.  As we have remarked, for the question of $\theta$ dependence of the free energy, one is interested in
much shorter distance scales.

%This is in accord with our discussion of the
%order by order coordinate
%space computation of the Greens function, we identified several classes of Feynman diagrams.  Here we would like to verify
%that, apart from one loop corrections with logarithmic sensitivity to $\mu$, all corrections are proportional to $g_3^2$.

 \subsection{Implications of the Debye Mass Calculation for the Topological Susceptibility}
 \label{implications}

 As we have remarked, one situation where it has been suggested that large corrections to the Debye mass might be important is in instanton computations of the free energy
 (topological susceptibility) at high temperatures\cite{borsanyinature,villadoro,sharma}.  It has been argued that the instanton computation is proportional to a large power of an infrared cutoff, and that this cutoff is the Debye length.  But as noted in \cite{dinedraperinstantons} the actual cutoff is $T$ (or $T^{-1}$ in coordinate space)\cite{gpy}.  As we have now seen, this is a regime where perturbation theory is valid, at least until one encounters magnetic divergences at high orders.
 These corrections are not likely to be numerically large, or terribly important for estimating, for example, the axion
 dark matter density\cite{dinedraperinstantons}.

\section{Infrared Sensitivity in the Instanton Computation}
\label{chicalculation}

While we have argued that the Debye mass is not the relevant cutoff for the instanton computation, we do expect
actual infrared divergences to appear at some order; in other words, we do not expect to be able to perform a semiclassical computation
of the topological susceptibility to arbitrary accuracy.
 In this section, after first considering the question of what does
play the role of the cutoff on the $\rho$ integration at large $\rho$, we turn to a determination of the order in perturbation
theory infrared divergences actually arise in the computation of the topological susceptibility.  In the spirit of our earlier 
Wilsonian analyses, we use this result to determine the irreducible uncertainty of the semiclassical computation.

\subsection{Instantons as a Perturbation at Large Distances}

At finite temperatures, for $r =\vert  \vec x \vert \gg \beta$,
the instanton takes the form:
$$A^{ia} ={ \epsilon_{aij} x^j \over (r^2 + r^3/(\rho^2 T))}
~~A^{4a} =-{  x^a \over (r^2 + r^3/(\rho^2 T))}
$$
%It has been said that the Debye mass plays the role of the infrared cutoff in the instanton computation.  Indeed, this
%is said somewhat heuristically in \cite{gpy}.  We have already mentioned that the effective cutoff is $T$, but it is interesting to see how this emerges from focussing on the effective three dimensional theory. 

From a three dimensional perspective, the instanton solution is well behaved at large distances, with $\vec E$ and $\vec B$
fields falling off as $1/r^2$, but singular at short distances.  The temperature and the scale size (which is of order $T$) act
as short distance cutoffs, yielding a finite action.  This is complicated to describe from our Wilsonian perspective.  In addition
to generating contributions to local operators, the short distance physics yields boundary conditions for
the three dimensional classical solutions as well as an integration measure.  Rather than describe the process of integrating
out short distance physics in this way, we will content ourselves with an examination of corrections to the leading
semiclassical approximation, isolating contributions which behave as $\log(T/m_{mag})$ and $\log(\rho m_{mag})$.  From
this analysis, we will infer the extent to which one can estimate $\chi$, and the irreducible uncertainty.

If we want to investigate actual infrared divergences, we need to study loop corrections to the instanton computation in this background.  Because
we are interested in effects at large distances, we are interested in integration regions where the fields of the instanton are small, and Greens
functions of the fluctuating fields are close to their free field expressions.  In particular, we can attempt to treat the instanton as a perturbation.  By this we mean that we break up the fields as
\beq
A_4 = A_{inst}^4 + a_4;~~~A^i = A_{inst}^i + a^i.
\label{fieldbreakup}
\eeq
Then there are interactions involving two fluctuating fields, $a^i$ proportional to one or two powers of the background field, and three powers of the fluctuating fields and one power of the background field.  For the instanton fields, we will take the large $r$ limits.  We will need
to integrate over collective coordinates for translations, dilitations and rotations.  The integral over $\rho$ will be controlled by the same
exponential terms as in the leading approximation, up to small corrections.  As a result, the dominant $\rho$ is of order $T^{-1}$.
The rotational collective coordinate is simple to deal with as $\chi$ is itself rotational and gauge invariant.

The translational
collective coordinate, $\vec x_0$, requires closer attention.    In perturbation theory at order
$n$, we will have vertices labeled by $x_i,~i = 1,\dots n$.  At vertices with insertions of the instanton field,
$A_{inst} = A_{inst}(\vec x_i - \vec x_0)$. We will also have products of free Greens functions (and derivatives),
$\Delta(x_i - x_j)$.  So if we shift $\vec x_i \rightarrow \vec x_i + \vec x_0$, the integral over $\vec x_0$ factors out, yielding
the factor of volume appropriate to the three dimensional vacuum energy.  The free propagators, in coordinate space, are
simply
\beq
\Delta(\vec x_i - \vec x_j) ={g_3^2 \over 4 \pi} {1 \over \vert \vec x_i - \vec x_j \vert}.
\eeq

\begin{figure}%[t!]
\includegraphics[width=10cm]{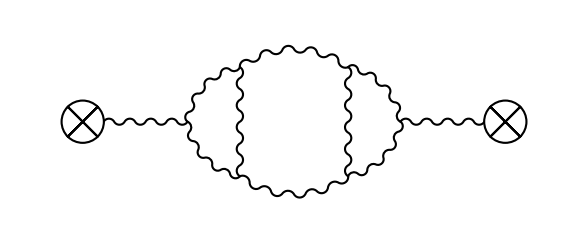}
\caption{Infrared divergent contributions to $\chi$ at three loops.} 
\label{threeloopchi}
\end{figure}

 The asymptotic behavior of the
instanton is:
\beq
(A_4)_{inst})^a = {\rho^2 T x^a\over x^3}~~~ (A^i_{inst})^a = {\rho^2 T \epsilon_{aij} x^j \over x^3}. 
\eeq
So formally, in the large distance regime we have an expansion in powers of $g_3^2$ and $\rho^\prime = \rho^2 T$.
We can power count on diagrams at $n+1$ loops.  These will have $n$ factors of $g_3^2$.  Then if there are $m$ insertions of the 
background field we have $m$ factors of $\rho^{\prime} = \rho^2 T$.   Schematically, the graph has the structure   
\beq
(g_3^2)^n \rho^{\prime~m} \prod \int d^3 x_i^{m+ 2n} \partial_i^{m+ 2n}{1 \over \vert {\vec x_i - \vec x_j}\vert^{3n + m}}
\eeq
where the partial derivatives are meant to indicate vertices with derivatives and the factors of $1 \over \vert \vec x_i - \vec x_j \vert$
to indicate the number of propagators.  Then we can assign a "superficial degree of divergence", $n-m$ to each graph.  Then if
\begin{enumerate}
\item  $n<m$, the graph has power law divergence in the ultraviolet, corresponding to domination of the contribution to the Wilsonian
action by high mo`menta.  Such diagrams will yield powers of $T$ relative to the leading contribution.
\item  $n=m$, the graph is logarithmically divergent in the ultraviolet and infrared, similar to the $g^6$ contributions to the 
perturbative free energy.
\item  $n>m$, the diagram exhibits a power law divergence in the infrared, and should be thought of as a contribution from
the low energy, three dimensional theory.
\end{enumerate}
For $n=m=1$, however, the relevant Feynman diagrams vanish.
So the action indicates infrared sensitivity first at order $g_3^4 \rho^{\prime 2}$ (figure \ref{threeloopchi}).

\subsection{Subtleties at Two Loops}

At two loop order, there are diagrams which, individually, are ultraviolet and infrared divergent.  These arise if, for example,
we consider an insertion of two instanton fields at a point, and integrate over the instanton pair.
 Graphically, this (and similar contributions) correspond to examining the effects of the instanton on the large distance behavior of the $A^i$ two-point function.  This effect can be summarized in terms of the insertion of a local operator.  Here it is necessary to consider 
\beq
\langle A^i(\vec z_1) A^j(\vec z_2) \rangle
\eeq
The corrections which arise from insertions of two powers of the background instanton field have the form::
\beq
\delta \langle A^i(\vec z_1) A^j(\vec z_2) \rangle =g_3^2 \int {d\rho \over \rho^4} \int d^3 x_0 f(\rho)\int d^3 z_3 {1 \over \vert \vec z_1 - \vec z_3 \vert}
{1 \over \vert \vec z_2 - \vec z_3 \vert}A_{inst}^2(\vec z_3 - \vec x_0).
\eeq
Shifting $z_3 \rightarrow z_3 + x_0$, if one estimates the integral over $z_3$ by ignoring the $z_3$ dependence of the first two propagators, one has:
\beq
\delta \langle A^i(\vec z_1) A^j(\vec z_2) \rangle =g_3^2\int {d\rho \over \rho^4} \int d^3 x_0 f(\rho) {1 \over \vert \vec z_1 - \vec x_0 \vert}
{1 \over \vert \vec z_2 - \vec x_0 \vert}\int d^3 z_3A_{inst}^2(\vec z_3).
\eeq
The $z_3$ integral is convergent and dominated by scales of order $\rho$.  Since $\rho \ll \vert \vec z_1 \vert,\vert \vec z_2 \vert$,
this is a short distance effect.  This is the result
one would obtain from a term in the effective lagrangian
\beq
\delta {\cal L} = \mu^2 (A^i)^2,
\eeq
with $\mu^2$ the result of the $z_3$ integral above.  But such a term is not gauge invariant, so the leading
short distance contributions must cancel.  The lowest dimension gauge invariant operator is
\beq
\delta{\cal L} = F_{ij}^2.
\eeq
This insertion of this operator at two loops does not lead to an expression which is infrared divergent.

So the leading divergence arises at three loops.  The form of the susceptibility is:
\beq
\chi(T) = \chi_0(T) (1 +a  g_3^2/T + b g_3^4 \log(T) + c g_3^4) 
\eeq
where $a$ and $b$ can be computed in the semiclassical approximation, but $c$ requires a lattice computation.  This last term
represents, again, the irreducible uncertainty in the semiclassical analysis.

\section{Conclusions}
\label{conclusions}

We have studied two physical quantities in high temperature QCD: the Debye mass and the topological susceptibility.  For $\chi$, we have seen that uncontrolled infrared divergences first arise at order $g^4$.  The leading semiclassical approximation would appear to be reliable at the fraction of a percent level.  

As we have reviewed, the Debye length -- which determines the exponential falloff of the $A_4$ Green's function (or the gauge invariant
Green's function which we have discussed) at very large distances -- is not critical to understanding the behavior of the susceptibility.  But it is interesting in its own right.  We have explained why, even though fundamentally a strong coupling problem, one can obtain a reliable estimate for this length.  This involves carefully considering the fact that the system is gapped and the structure of the 
perturbation series.  We have seen that, as a Minkowski theory, one can calculate
the leading order contribution to the position of the pole, which is larger by a logarithm then the expected uncertainties.
Infrared divergences appear in this computation at precisely the order where one expects confinement effects to be important.
We have stressed, as in \cite{dinedraperinstantons}, that these large corrections to the Debye mass are not important for the calculation
of the susceptibility.
%We have also seen that, in practice, the perturbative corrections to the Debye length are not numerically much different than those obtained by evaluating the propagator on the lowest order pole.

\vskip 1cm
\noindent
%{\bf Acknowledgements:} 
\noindent
{\bf Acknowledgements:}  This work was supported in part by the U.S. Department of Energy grant number DE-FG02-04ER41286.  We would like to thank Tom Banks and Patrick Draper for conversations, and Peter Arnold and Larry Yaffe for their critical reading of an early version of the manuscript.

%\bibliography{debye_mass}{}
%\bibliographystyle{utphys}
\bibliography{debye_length_qcd.bbl}

\providecommand{\href}[2]{#2}\begingroup\raggedright\begin{thebibliography}{10}

\bibitem{braatenpisarski}
E.~Braaten and R.~D. Pisarski, ``{Simple effective Lagrangian for hard thermal
  loops},''
\href{http://dx.doi.org/10.1103/PhysRevD.45.R1827}{{\em Phys. Rev.} {\bfseries
  D45} no.~6, (1992) R1827}.
%%CITATION = PHRVA,D45,R1827;%%.

\bibitem{rebhan}
A.~K. Rebhan, ``{The NonAbelian Debye mass at next-to-leading order},''
  \href{http://dx.doi.org/10.1103/PhysRevD.48.R3967}{{\em Phys. Rev.}
  {\bfseries D48} (1993) R3967--R3970},
\href{http://arxiv.org/abs/hep-ph/9308232}{{\ttfamily arXiv:hep-ph/9308232
  [hep-ph]}}.
%%CITATION = HEP-PH/9308232;%%.

\bibitem{arnoldyaffe}
P.~B. Arnold and L.~G. Yaffe, ``{The NonAbelian Debye screening length beyond
  leading order},'' \href{http://dx.doi.org/10.1103/PhysRevD.52.7208}{{\em
  Phys. Rev.} {\bfseries D52} (1995) 7208--7219},
\href{http://arxiv.org/abs/hep-ph/9508280}{{\ttfamily arXiv:hep-ph/9508280
  [hep-ph]}}.
%%CITATION = HEP-PH/9508280;%%.

\bibitem{gpy}
D.~J. Gross, R.~D. Pisarski, and L.~G. Yaffe, ``{QCD and Instantons at Finite
  Temperature},''
\href{http://dx.doi.org/10.1103/RevModPhys.53.43}{{\em Rev. Mod. Phys.}
  {\bfseries 53} (1981) 43}.
%%CITATION = RMPHA,53,43;%%.

\bibitem{bonatia}
C.~Bonati, M.~D'Elia, M.~Mariti, G.~Martinelli, M.~Mesiti, F.~Negro,
  F.~Sanfilippo, and G.~Villadoro, ``{Axion phenomenology and
  $\theta$-dependence from $N_f = 2+1$ lattice QCD},''
  \href{http://dx.doi.org/10.1007/JHEP03(2016)155}{{\em JHEP} {\bfseries 03}
  (2016) 155},
\href{http://arxiv.org/abs/1512.06746}{{\ttfamily arXiv:1512.06746 [hep-lat]}}.
%%CITATION = ARXIV:1512.06746;%%.

\bibitem{bonatib}
C.~Bonati, M.~D'Elia, M.~Mariti, G.~Martinelli, M.~Mesiti, F.~Negro,
  F.~Sanfilippo, and G.~Villadoro, ``{Recent progress on QCD inputs for axion
  phenomenology},'' \href{http://dx.doi.org/10.1051/epjconf/201713708004}{{\em
  EPJ Web Conf.} {\bfseries 137} (2017) 08004},
\href{http://arxiv.org/abs/1612.06269}{{\ttfamily arXiv:1612.06269 [hep-lat]}}.
%%CITATION = ARXIV:1612.06269;%%.

\bibitem{borsanyia}
S.~Borsanyi, M.~Dierigl, Z.~Fodor, S.~D. Katz, S.~W. Mages, D.~Nogradi,
  J.~Redondo, A.~Ringwald, and K.~K. Szabo, ``{Axion cosmology, lattice QCD and
  the dilute instanton gas},''
  \href{http://dx.doi.org/10.1016/j.physletb.2015.11.020}{{\em Phys. Lett.}
  {\bfseries B752} (2016) 175--181},
\href{http://arxiv.org/abs/1508.06917}{{\ttfamily arXiv:1508.06917 [hep-lat]}}.
%%CITATION = ARXIV:1508.06917;%%.

\bibitem{borsanyinature}
S.~Borsanyi {\em et al.}, ``{Calculation of the axion mass based on
  high-temperature lattice quantum chromodynamics},''
  \href{http://dx.doi.org/10.1038/nature20115}{{\em Nature} {\bfseries 539}
  no.~7627, (2016) 69--71},
\href{http://arxiv.org/abs/1606.07494}{{\ttfamily arXiv:1606.07494 [hep-lat]}}.
%%CITATION = ARXIV:1606.07494;%%.

\bibitem{villadoro}
G.~Grilli~di Cortona, E.~Hardy, J.~Pardo~Vega, and G.~Villadoro, ``{The QCD
  axion, precisely},'' \href{http://dx.doi.org/10.1007/JHEP01(2016)034}{{\em
  JHEP} {\bfseries 01} (2016) 034},
\href{http://arxiv.org/abs/1511.02867}{{\ttfamily arXiv:1511.02867 [hep-ph]}}.
%%CITATION = ARXIV:1511.02867;%%.

\bibitem{dinedraperinstantons}
M.~Dine, P.~Draper, L.~Stephenson-Haskins, and D.~Xu, ``{Axions, Instantons,
  and the Lattice},'' \href{http://dx.doi.org/10.1103/PhysRevD.96.095001}{{\em
  Phys. Rev.} {\bfseries D96} no.~9, (2017) 095001},
\href{http://arxiv.org/abs/1705.00676}{{\ttfamily arXiv:1705.00676 [hep-ph]}}.
%%CITATION = ARXIV:1705.00676;%%.

\bibitem{kisslinger}
P.~D. Morley and M.~B. Kislinger, ``{Relativistic Many Body Theory, Quantum
  Chromodynamics and Neutron Stars/Supernova},''
\href{http://dx.doi.org/10.1016/0370-1573(79)90005-X}{{\em Phys. Rept.}
  {\bfseries 51} (1979) 63}.
%%CITATION = PRPLC,51,63;%%.

\bibitem{schwartzfieldtheory}
M.~D. Schwartz, {\em {Quantum Field Theory and the Standard Model}}.
\newblock Cambridge University Press, 2014.
\newblock
\url{http://www.cambridge.org/us/academic/subjects/physics/theoretical-physics-and-mathematical-physics/quantum-field-theory-and-standard-model}.
\newblock
%%CITATION = INSPIRE-1276589;%%.

\bibitem{sharma}
P.~Petreczky, H.-P. Schadler, and S.~Sharma, ``{The topological susceptibility
  in finite temperature QCD and axion cosmology},''
  \href{http://dx.doi.org/10.1016/j.physletb.2016.09.063}{{\em Phys. Lett.}
  {\bfseries B762} (2016) 498--505},
\href{http://arxiv.org/abs/1606.03145}{{\ttfamily arXiv:1606.03145 [hep-lat]}}.
%%CITATION = ARXIV:1606.03145;%%.

\end{thebibliography}\endgroup
%\bibliographystyle{unsrt}
%\bibliographystyle{JHEP}
%\bibliography{Biblio}

\end{document}